# Observation of Bloch oscillations in complex PT-symmetric photonic lattices

Martin Wimmer[1,2], Mohammed-Ali Miri[3], Demetrios Christodoulides[3], *Ulf Peschel[4]

[1]Institute of Optics, Information and Photonics, Friedrich-Alexander-Universität Erlangen-Nürnberg, Staudtstraße 7/B2, 91058 Erlangen, Germany

[2]Erlangen Graduate School in Advanced Optical Technologies (SAOT), 91058 Erlangen, Germany

[3]CREOL, College of Optics and Photonics, University of Central Florida, Orlando, Florida 32816–2700, USA

[4]Institute of Solid State Theory and Optics, Friedrich Schiller University Jena, Max-Wien-Platz 1, 07743 Jena

**Abstract:**

Light propagation in periodic environments is often associated with a number of interesting and potentially useful processes. If a crystalline optical potential is also linearly ramped, light can undergo periodic Bloch oscillations, a direct outcome of localized Wannier-Stark states and their equidistant eigenvalue spectrum. Even though these effects have been extensively explored in conservative settings, this is by no means the case in non-Hermitian photonic lattices encompassing both amplification and attenuation. Quite recently, Bloch oscillations have been predicted in parity-time-symmetric structures involving gain and loss in a balanced fashion. While in a complex bulk medium, one intuitively expects that light will typically follow the path of highest amplification, in a periodic system this behavior can be substantially altered by the underlying band structure. Here, we report the first experimental observation of Bloch oscillations in parity-time-symmetric mesh lattices. We show that these revivals exhibit unusual properties like secondary emissions and resonant restoration of PT symmetry. In addition, we present a versatile method for reconstructing the real and imaginary components of the band structure by directly monitoring the light evolution during a cycle of these oscillations.

Introduction

Bloch oscillations were first considered within the context of solid state physics by Bloch and Zener[1,2]. While a free electron under the influence of a constant electric field experiences uniform acceleration, the situation is completely different when this same charge carrier is placed in a crystal. Interestingly, in this latter setting, the transport behavior is directly governed by the band structure of the periodic medium, which imposes a cyclic motion to the particle, better known as a Bloch oscillation. Bloch revivals represent a general phenomenon that can be observed in many and diverse physical settings like ultra-cold atoms[3], optical waveguide arrays[4] and in semiconductor superlattices[5]. Given the nature of these latter platforms, Bloch oscillations were thus far studied in Hermitian or conservative arrangements. Recently, however, the prospect of observing this class of oscillations in non-Hermitian PT-symmetric optical arrangements has been proposed[6,7]. As shown in Refs. [6,7], in this case, Bloch oscillations are expected to display unusual features such as non-reciprocal cycles related to a violation of Friedel's law of Bragg scattering and cascade of wavepacket splittings.

In photonics, even under the best conditions, attenuation is omnipresent and hence amplifiers are often used to compensate for loss. As recently indicated in a number of studies, this non-Hermitian behavior can be appropriately tailored by introducing parity-time (PT) symmetry[8] in the system itself. Within the framework of optics, this symmetry can be established by judiciously incorporating gain and loss in a balanced fashion[9]. In this respect, the real part of the refractive index distribution must have a symmetric profile whereas the imaginary component must be antisymmetric. PT-symmetric optical arrangements are known to exhibit a number of unexpected properties and characteristics including power unfolding[9], abrupt phase transitions[10-12], breaking of left-right symmetry[12], simultaneous lasing-absorbing[13,14], selective mode lasing[15-17], and unidirectional invisibility[18-20].

In a system encompassing amplifying and attenuating regions, one may anticipate that light will follow the direction of gain. Yet, as we will see, Bloch oscillations in the presence of both gain and loss provide a counter example refuting this perception. Thus, the question arises, as to which process will ultimately dominate the other. Will light still follow the direction of amplification, or will Bloch oscillations instead force the wavepacket to enter the loss region and hence experience attenuation?

In this Report, we address this intriguing question, by considering the role of the complex band structure on Bloch oscillations when taking place in non-Hermitian PT-symmetric mesh lattices. We show that by controlling the period of Bloch oscillations, one can establish pseudo-Hermitian wavepacket propagation that continuously undergoes through successive cycles of amplification and attenuation. Along these lines we observe a resonant restoration of PT symmetry as well as secondary emissions. This special case of PT symmetric Bloch oscillations is then generalized to arbitrary Bloch gradients in locally PT-symmetric systems. In this case, the wavepacket trajectories in this mesh lattice can be analyzed in order to reconstruct both the real and imaginary parts of the band diagram associated with this system.

**Experimental setup and time multiplexing**

In what follows, we provide a brief introduction into time multiplexed systems based on coupled fiber loops and their relation to mesh lattices. In our experiments, we use two coupled fiber loops of different lengths, to emulate a 1+1D lattice (see Fig. 1a)[12]. Erbium doped fiber amplifiers compensate losses inside the fiber loops and additionally, acousto-optical modulators are inserted and set to a transmission of 50%. By increasing or decreasing this transmission ratio, an effective gain or loss is generated. A detailed explanation of the experiment is reported in [21]. Depending on whether pulses travel through the longer or shorter loop, they either propagate on the equivalent lattice from "North West" to "South East" or vice versa (see Fig. 1b). Thus, their position on the lattice is encoded by the arrival times after $m$ roundtrips[22]. Note that in each roundtrip, a pulse can jump either to an earlier or later arrival time slot, thus decreasing or increasing its position $n$ on the equivalent lattice by $\pm 1$. In the experiment, we start with a single pulse at time $m = 0$ and at position $n = 0$. After passing through the 50/50 coupler the pulse splits up into two pulses at time $m = 1$ and at position $n = -1$ and $n = 1$ with half of the intensity. Furthermore, the fiber coupler induces a phase shift of $\pi/2$ for the case of a crossover pulse.

To mathematically describe this process, pulses passing through the short/long loop and therefore arriving earlier/later or traveling to the West/East in the equivalent spatial scheme (see Fig.1b) are labeled as $u_n^m$ / $v_n^m$, respectively. As dispersive spreading of the pulses is negligible only their amplitudes are of interest, while the pulse shape itself is neglected. In this setting, the evolution of an initial distribution of pulse amplitudes $u_n^0$ and $v_n^0$ can be described by iteratively applying the evolution equations[12]

$$u_n^{m+1} = \frac{\sqrt{G_u}}{\sqrt{2}}(u_{n+1}^m + iv_{n+1}^m)e^{i\varphi(m)}, \qquad (1.1)$$

$$v_n^{m+1} = \frac{\sqrt{G_v}}{\sqrt{2}}(v_{n-1}^m + iu_{n-1}^m), \qquad (1.2)$$

where $G_u$ and $G_v$ are the gain and loss parameters for pulses inside the short (long) loop which travel on the lattice toward the East (West). In addition, a phase modulator is inserted into the shorter loop, which allows an arbitrary modulation of the phases of the pulses by inducing a time dependent shift $\varphi(m)$.

This technique, which is referred to as time multiplexing[22], does not only drastically reduce the number of components, since each element in the experiment is involved in every single round trip, but it also provides an extremely stable measurement platform. Although pulses from a non-temperature-stabilized DFB laser diode propagate for up to 1600 km corresponding to 400 roundtrips, they can still interfere with each other with high contrast. This provides an additional advantage given that in the two coupled fiber loops, only pulses which have traveled for the same time through the longer and shorter loop can interfere. Therefore, any low frequency noise is not involved in our measurements since all pulses are affected in exactly the same way. In principle, the setup is equivalent to a self-adjusting interferometer.

**Derivation of the band structure**

To better understand our system, we will first focus our discussion on the equivalent spatial mesh lattice, depicted in Fig. 1b. In this arrangement, the Floquet-Bloch mode eigenstates can be described according to:

$$\begin{pmatrix} u_n^m \\ v_n^m \end{pmatrix} = \begin{pmatrix} U \\ V \end{pmatrix} e^{i\theta m} e^{iQn} \qquad (2)$$

where $\theta$ is the longitudinal propagation constants and $Q$ is the transverse wavenumber, also referred to as Bloch momentum. In our system, the eigenmodes $(U,V)^t$ implicitly contain information as to the amplitude and phase relation between the longer and shorter loop. The band structure or dispersion relation of this system is given by[23]

$$\cos\theta = \frac{1}{\sqrt{2}}\cos Q, \qquad (3)$$

(derived by inserting Eq. 2 into Eq. 1). Evidently it resembles a set of coupled two level systems, with an upper and a lower band separated by a gap (see Fig. 2a). A specific point inside the band structure can be excited by launching a broad Gaussian distribution in both loops, and by appropriately tuning the phase of the pulses in the shorter loop in order to match the state $(U,V)^t$.

**Hermitian Bloch oscillations**

In our platform, applying a constant phase gradient $\varphi(m,n)$ either along the temporal (vertical) or the spatial (horizontal) direction is similar to a constant electric field applied to a crystal. Thus, if a single pulse is injected into the fiber system in presence of a phase modulation $\varphi(m) = m\varphi_0$, the entire Brillouin zone is excited, thus resulting in a Bloch oscillation of the wavepacket (see Fig 2b). In contrast, spectrally resolved measurements require the use of broad initial distributions, which only excite a narrow region within the Brillouin zone. Hence, starting with a Gaussian profile, one can approximately select a specific value of the Bloch momentum $Q_0$. In the presence of a phase gradient, at each time step, the Bloch momentum $Q$ is shifted by a small step, i.e.:

$$Q = Q_0 + m\frac{\varphi_0}{2}. \qquad (4)$$

The cyclic motion through the Brillouin zone makes the wavepacket undergo Hermitian Bloch oscillations in real space (see Fig. 2c).

**Bloch oscillations in globally PT-symmetric lattices**

Introducing amplification and attenuation into the system in a balanced way $G = G_u = G_v^{-1}$ and by switching the gain and loss after every roundtrip leads to an equivalent photonic lattice with vertical gain and loss channels that are coupled at discrete time steps $m$ (see Fig. 3a). In order to establish full PT symmetry, one would also have to apply a periodic phase modulation to the system[12]. In the absence of a periodic phase modulation, however, the system enters the broken PT symmetry regime and as a result a part of the band structure is complex, with the exceptional points marking the transition between the real and complex band regions (see Fig. 3b).

The idiosyncrasies of this lattice are also reflected in our measurements, where we again start with a broad Gaussian distribution in order to probe the spectral features of the band structure. In this regard, we find that the wave packet is strongly amplified in the regions of the band structure, where the eigenvalues have a non-vanishing imaginary part. Furthermore, the exceptional points lead to the formation of secondary emissions which emerge each time the wave packet passes through the exceptional point (see Fig. 4).

While in general, the existence of complex eigenvalues introduces exponential growth during Bloch oscillations, there are specific resonant values of the Bloch gradient[6] which result in a pseudo-Hermitian motion with a vanishing net increase in power. In this case, a balanced passage along branches of the band structure with positive and negative imaginary eigenvalues causes the energy of the wave packet to periodically increase and decrease while the average total power remains constant over two Bloch oscillations (see Fig. 4). In other words, PT symmetry is restored at these resonant values. This process is elucidated in Fig. 4. As predicted in [6] we found the reciprocal values of these resonant Bloch gradients to have an equidistant spacing defined by the amplification value.

**Local PT symmetry**

Another interesting non-Hermitian platform where Bloch oscillations can be observed and studied is that of local PT-symmetric mesh lattices[21]. Such system can be readily implemented by time multiplexing techniques where all pulses inside the long loop are amplified, while at the same time the pulses inside the short loop are attenuated (see Fig. 5a)[21]. Figure 5a depicts the spatial analogue of this lattice that happens to be transversely PT-symmetric at every longitudinal step $m = m_0$ (locally). Using the same ansatz as in Eq. 2, one can directly find the band structure of this new arrangement which is given by:

$$\cos\theta = \frac{1}{\sqrt{2}}\cos\left(Q - \frac{i}{2}\gamma\right), \qquad (5)$$

where $\gamma = \log(G)$. In this case, the propagation constant $\theta$ is complex for all values of the Bloch momentum $Q$, except at the center and the edges of the Brillouin zone ($Q = 0$ and $Q = \pm\pi$) (see Fig. 5b). Hence, light propagation in this lattice is by nature non-Hermitian and following a naive interpretation one would always expect exponential growth.

Still, one may ask the question: Will light follow the direction of gain by always propagating in the gain loop or will it perform Bloch oscillations, if a linear phase gradient $\varphi_0$ is applied? Interestingly, the band structure still dictates the dynamics and as a result, Bloch oscillations take place in real space in the conventional way, thus disregarding the presence of gain and loss channels (see Figs. 5c and 5d). Because light follows one of the bands and as there are no exceptional points to switch between the bands, the field distribution is periodically amplified and attenuated during these Bloch oscillations (see Fig. 5e) but the average power stays constant.

Integrating the imaginary part of the dispersion relation (see Fig. 5b) from $Q = -\pi$ to $Q = \pi$ yields a vanishing imaginary component which explains why after one cycle of Bloch oscillations no overall increase in the power of the wave packet is observed (see Fig. 5e). Thus, similar to common PT-symmetric systems, the introduction of a phase potential can again lead to a pseudo-Hermitian propagation.

Based on the previous observations, we can now establish a procedure through which we can experimentally reconstruct the complex band structure. To do so, we monitor at each time step the center of mass $n_0(m)$ (see Fig. 5d) and the intensity $I_0(m)$ (see Fig. 5e) of the wavepacket by using a Gaussian fit. By taking temporal derivatives, the group velocity $v_G(m)$ and the change in power $p_0(m)$ can be determined:

$$v_G(m) = \frac{\partial}{\partial m}n_0(m) \qquad (6)$$

$$p_0(m) = \frac{\partial}{\partial m}I_0(m). \qquad (7)$$

Since the phase potential shifts the Bloch momentum $Q$ by $\varphi_0/2$ at every step, after $m$ rounds the Bloch momentum becomes equal to $Q = Q_0 + m\varphi_0/2$. For experimental convenience, the initial wave packet is injected at the center of the Brillouin zone at $Q_0 = 0$. The imaginary part of the band structure can be evaluated from the growth/decay rate,

$$\text{Im}[\theta(Q)] = -\frac{1}{2}\frac{\partial}{\partial m}\log[I_0(m)]. \qquad (8)$$

The corresponding real part of the dispersion relation is then determined (up to an unknown constant) by integrating the group velocity

$$\text{Re}[\theta(Q)] = \text{Re}[\theta(Q = -\pi)] + \int_{Q'=-\pi}^{Q} v_G(Q')dQ' \qquad (9)$$

Thus, by evaluating Eqs. 8 and 9, the entire complex band structure can be reconstructed as shown in Fig. 5e. The experimentally obtained reconstruction (see Fig. 5f) is in excellent agreement with the theoretically calculated dispersion relation (see Fig. 5b).

**Conclusions**

In conclusion, we have experimentally demonstrated optical Bloch oscillations in global and local PT-symmetric mesh lattices. Our studies indicate, that during these oscillations, this class of non-Hermitian systems can effectively behave in a pseudo-conservative fashion. In the case of global PT-symmetric systems, stable wave propagation was observed for a specific set of resonant Bloch gradients. For other values, secondary emissions occur during growing Bloch revivals. On the other hand, in locally PT-symmetric arrangements, the propagation was found to be always stable due to a cyclic amplification and attenuation process. In addition we have shown, that Bloch oscillations can be used to reconstruct the whole bandstructure including complex eigenvalues. Given the universality of Bloch oscillations, our results may be pertinent to other areas of physics where non-Hermiticity plays a role in a periodic environment.

**Acknowledgements**

We acknowledge financial support from DFG Forschergruppe 760, DFG Project PE 523/10-1, Cluster of Excellence Engineering of Advanced Materials (EAM) and School of Advanced Optical Technologies (SAOT). This work was also supported by NSF (grant ECCS-1128520), AFOSR (grants FA9550-12-1-0148 and FA9550-14-1-0037).

**Additional information**

Correspondence and requests for materials should be addressed to U.P. (email: ulf.peschel@uni-jena.de)

**Author contributions**

M.W. performed the measurements. M.W., M-A.M., D.N.C. and U.P. developed the theoretical background.

**Competing financial interests**

The authors declare no competing financial interests.

**Figures**

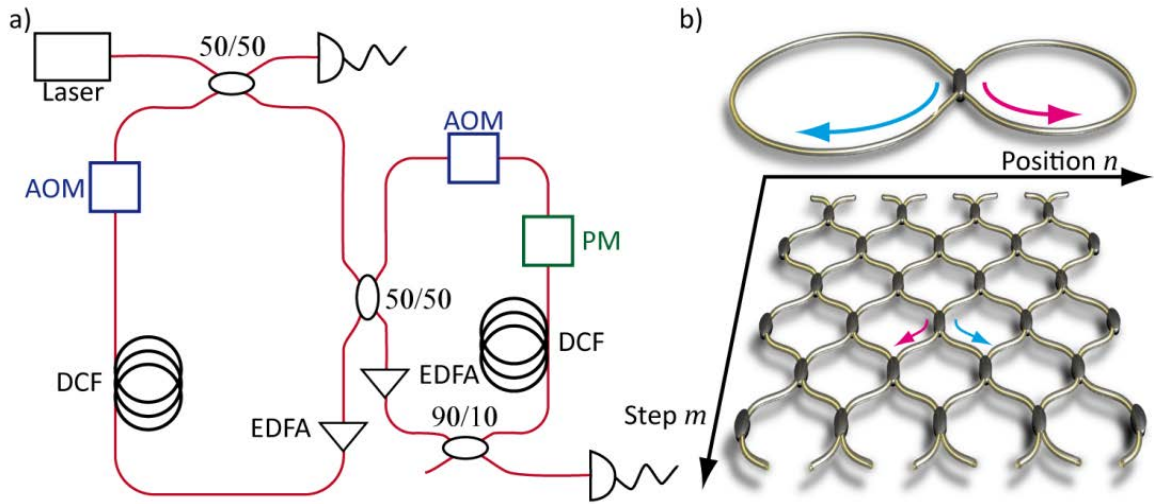

**Figure 1 | Experimental implementation of time multiplexing for generating 1+1D mesh lattices. a**, Two fiber loops of different lengths are connected by a 50/50 fiber coupler. Acousto-optical modulators (AOM) are used for generating an effective gain and loss distribution. The Bloch gradient is created by a phase modulator (PM) inside the short loop. Erbium doped fiber amplifiers (EDFA) are used in both loops to compensate for any unwanted losses. **b,** A roundtrip inside the long (short) loop is equivalent to propagating from North West (North East) to South East (South West) on the lattice.

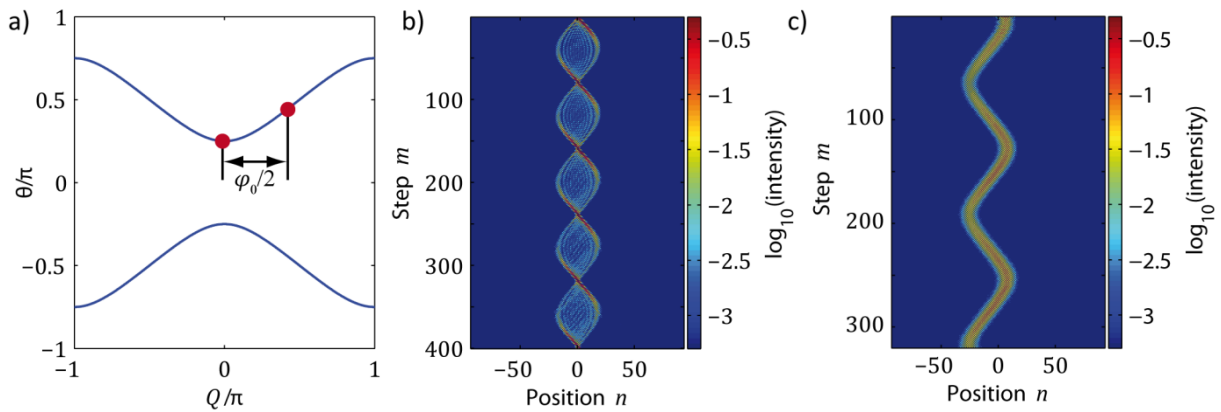

**Figure 2 | Experimental observation of Bloch oscillations in a conservative system. a,** The band structure of the conservative mesh lattice consists of two bands separated by a band gap. The Bloch gradient $\varphi_0$ along the temporal direction increases the Bloch momentum (starting from $Q_0 \approx 0$) of an initial excitation (red circle) at each time step by $\frac{\varphi_0}{2}$. **b,** If only a single lattice side is populated by the initial pulse distribution, Bloch oscillations of the entire envelope take place, having a periodicity of $T = 4\pi/\varphi_0$ (here $\varphi_0 \approx \pi/20$). **c,** In the case of a broad Gaussian initial excitation, the wave packet performs an oscillatory motion, as the state is shifted periodically through the Brillouin zone ($\varphi_0 \approx \pi/32$).

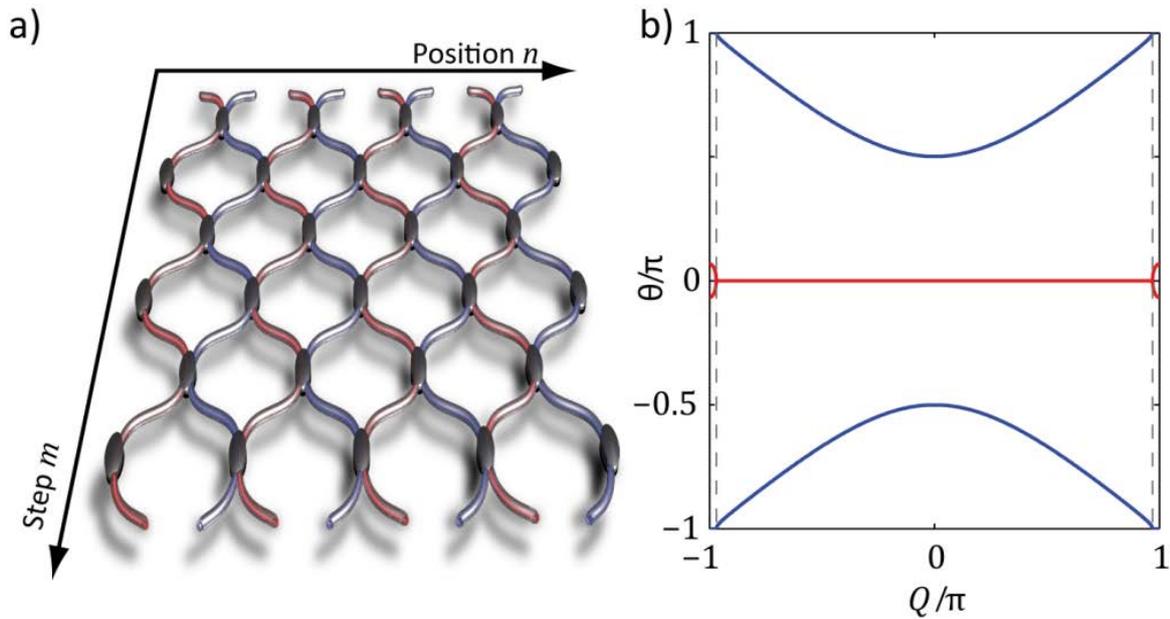

**Figure 3 | PT mesh lattice with a broken PT symmetry. a,** By alternating gain and loss in every other round trip, a 1+1D mesh lattice is established with vertical stripes of gain and loss. Due to the absence of a symmetric phase modulation, the PT phase is broken and exceptional points separating the conservative band structure from regions with complex eigenvalues appear (**b**).

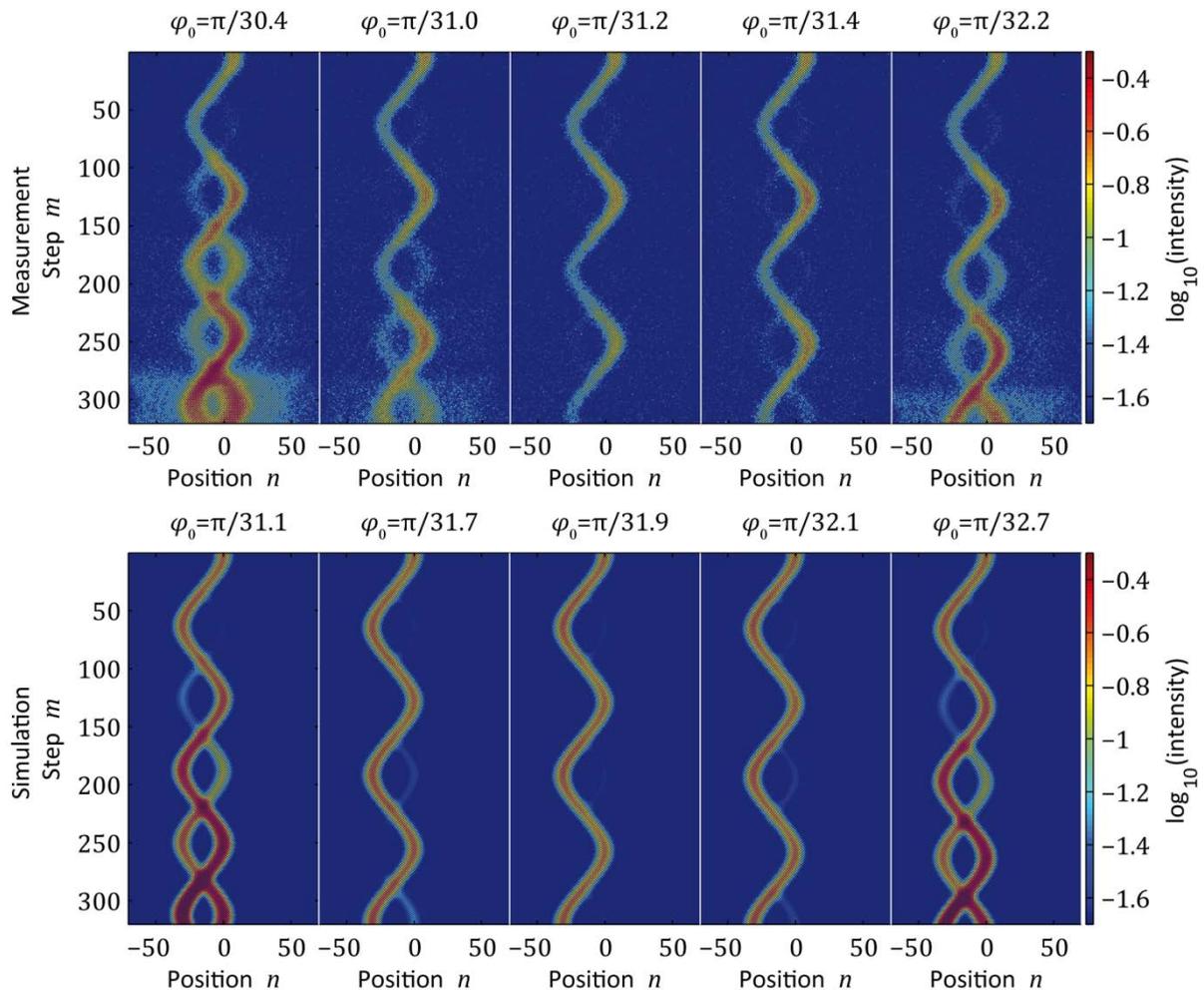

**Figure 4 | Experimental and numerical observation of Bloch oscillations in a system with a broken PT symmetry.** In all cases, the initial distribution is a Gaussian wavepacket occupying the upper band. Due to the partially complex band structure, the wavepacket is periodically amplified during Bloch oscillations. Note, that each time, the wavepacket passes through an exceptional point, a new branch of Bloch oscillations appears (secondary emissions). However, for resonant Bloch gradients ($\varphi_0 \approx \pi/31.2$ in the experiment and $\pi/31.9$ in our simulation), a pseudo-Hermitian propagation is restored, where the total energy stays constant over many time steps. Here, the gain/loss coefficient is $G = 1.1$.

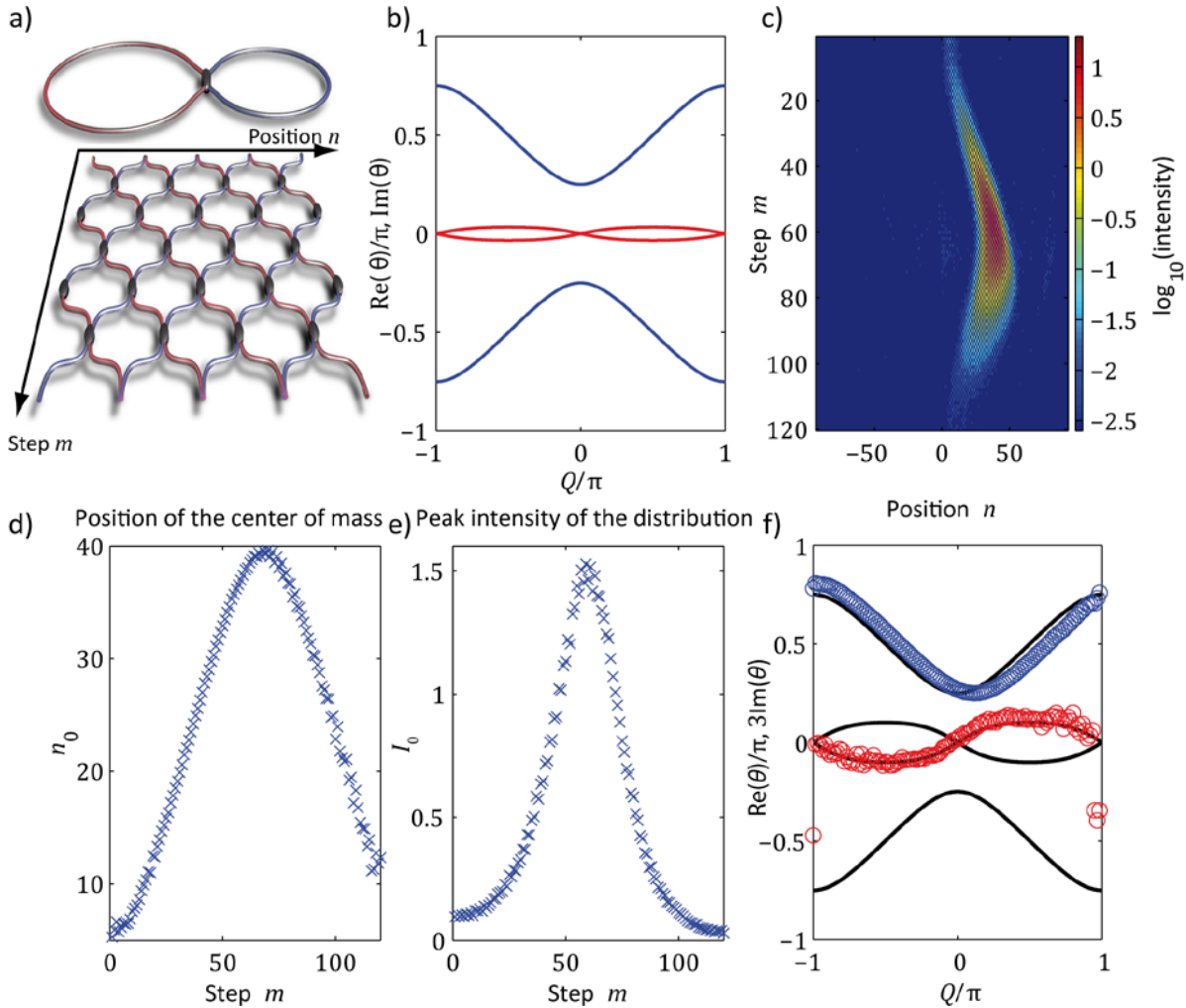

**Figure 5 | Experimental reconstruction of the band structure of a locally PT-symmetric mesh system by analyzing Bloch oscillations. a,** To implement a locally PT lattice, only pulses in the long loop are amplified, while the pulses in the short loop are attenuated. **b,** the resulting band structure is complex over the entire Brillouin zone (blue and red correspond to real and imaginary part). **c,** Observed Bloch oscillations for $\varphi_0 \approx \pi/30$. **d,** Estimated position of the wavepacket's center of mass. In this figure, the horizontal axis is mapped onto the Bloch momentum $Q$ by evaluating Eq. 4. **e,** The peak power of the Gaussian packet depends on the position within the Brillouin zone during Bloch oscillations. **f,** Reconstructed band structure of this locally PT-symmetric lattice (imaginary part is magnified by a factor of 3 for better visibility).